\documentclass[aps,prl,fleqn,twocolumn,superscriptaddress,floatfix]{revtex4}

\usepackage{epsfig}
\usepackage{amsmath,amssymb,amsfonts}

\begin{document}

\title{Thermal Recombination: Beyond the Valence Quark Approximation}

\author{B.~M\"uller}
\affiliation{Department of Physics, Duke University, Durham, NC 27708}

\author{R.~J.~Fries}
\affiliation{School of Physics and Astronomy, University of Minnesota,
             Minneapolis, MN 55455}

\author{S.~A.~Bass}
\affiliation{Department of Physics, Duke University, Durham, NC 27708}
\affiliation{RIKEN BNL Research Center, Brookhaven National Laboratory, 
             Upton, NY 11973}

\date{\today}
\preprint{NUC-MINN-05/4-T}

\begin{abstract}
Quark counting rules derived from recombination models agree 
well with data on hadron production at intermediate transverse 
momenta in relativistic heavy-ion collisions. They convey a
simple picture of hadrons consisting only of valence quarks.
We discuss the inclusion of higher Fock states that add sea
quarks and gluons to the hadron structure. We show that, 
when recombination occurs from a thermal medium, hadron 
spectra remain unaffected by the inclusion of higher Fock states.
However, the quark number scaling for elliptic flow is somewhat 
affected. We discuss the implications for our understanding of 
data from the Relativistic Heavy Ion Collider.
\end{abstract}

\maketitle

The production of hadrons with transverse momenta of a few GeV/c in 
Au+Au collisions at the Relativistic Heavy Ion Collider (RHIC) has
been found to exhibit certain unusual features. The emission of
baryons, in comparison with mesons, is significantly enhanced. 
For example, the ratio of protons to pions is about three times 
larger than in $p+p$ collisions \cite{phenix:03ppi}. Furthermore, 
the nuclear suppression factor $R_{AA}$ below 4 GeV/$c$ is close 
to unity for protons and $\Lambda$-hyperons, 
while it is about 0.3 for pions \cite{Velkovska:2004cb}. 
The azimuthal anisotropy
(called ``elliptic flow'' $v_2$)  of baryons lags behind 
that of mesons at low transverse momenta but exceeds the anisotropy of 
meson emission above 2 GeV/c \cite{v2_data,Sorensen:2003wi}. The 
recombination of quasi-thermal, deconfined quarks has been proposed 
as an explanation for these peculiarities 
\cite{Vo02,Fr03a,F03prc,Fr04,Gr03a,MV03,Hw03a}. 

These recombination models are based on the concept of {\em constituent 
quark recombination}, which assumes that the probability for the emission 
of a hadron from a deconfined medium is proportional to the probability 
for finding the valence quarks of the hadron in the density matrix 
describing the source. The baryon enhancement, as well as the different 
momentum dependence of meson and baryon anisotropies, rely essentially
on the different number of valence quarks in mesons (two) and baryons
(three). The simplicity of this concept has been criticized, because 
it does not do justice to the complexity of the internal structure of
hadrons in quantum chromodynamics (QCD). It is the purpose of this
article to address some of these objections by showing that important 
features of the emission of relativistic hadrons by sudden recombination 
from a \emph{thermal} deconfined medium do not depend on specific assumptions 
about their internal structure and survive in a more complete description 
of the emitted hadrons.

The quark-gluon content of a hadron in QCD, as that of a bound state
in any strongly interacting quantum field theory, is a function of
the resolution scale $Q^2$. The larger the scale $Q^2$ is, the more 
quantum fluctuations can be seen, and the number of gluons and sea
quarks in a hadron increases. Moreover, the structure of a hadron
depends on the frame of reference. This is so because the boost operator
$\mathbf K$ and the Hamiltonian $H$ generally do not commute: 
$[{\mathbf K},H]\ne 0$. Furthermore the quark-gluon structure depends
on the measurement process, i.~e.\ it is not universal. Different
processes probe the Fock states in a hadron with different weights.

In the following we will use a light-cone frame, where formally the
hadron momentum $P \to \infty$ and the momentum fractions of the partons
are the only dynamic degrees of freedom. A meson $M$ with valence quarks
$q_\alpha$ and $\bar q_\beta$ can then be written as an expansion in 
terms of increasingly complex Fock states:
\begin{widetext}
\begin{eqnarray}
|M\rangle 
&=& \int_0^1 dx_a dx_b \delta(x_a+x_b-1) 
          c_1(x_a,x_b) \left|q_\alpha(x_a){\bar q}_\beta(x_b)\right\rangle 
\nonumber \\
&+& \int_0^1 dx_a dx_b dx_c \delta(x_a+x_b+x_c-1) 
	  c_2(x_a,x_b,x_c) \left|q_\alpha(x_a){\bar q}_\beta(x_b)
	  g(x_c)\right\rangle 
\\
&+& \int_0^1 dx_a dx_b dx_c dx_d \delta(x_a+x_b+x_c+x_d-1) 
          c_3(x_a,x_b,x_c,x_d) 
          \left|q_\alpha(x_a){\bar q}_\beta(x_b)q_\gamma(x_c)
		 {\bar q}_\gamma(x_d)\right\rangle + \ldots
\nonumber 
\label{eq01}
\end{eqnarray}
\end{widetext}
Hence $x_i {\mathbf P}$ is the momentum of parton $i$. We have suppressed
the momentum $\mathbf P$ in the notation for 
simplicity. For convenience, we adopt a probabilistic normalization 
for partons: $\langle q_\alpha(x_a) | q_\beta(x_b)\rangle
=\delta_{\alpha\beta}\delta(x_a-x_b)$.
This implies that the completeness relation for the Fock space expansion 
(\ref{eq01}) takes the simple form:
\begin{eqnarray}
1 &=& C_1 + C_2 + C_3 + \cdots \nonumber \\
&\equiv& \int_0^1 dx_a dx_b\, \delta(x_a+x_b-1)\, 
\left|c_1(x_a,x_b)\right|^2
\\
&& + \int_0^1 dx_a dx_b dx_c\, \delta\left(\sum_i x_i -1\right) 
             \left|c_2(x_a,x_b,x_c)\right|^2 
\nonumber \\ &&+ \ldots \nonumber
\label{eq02}
\end{eqnarray}
A pure valence quark description would only include the 
first term, without accounting for its $Q^2$-dependence, which is
governed by renormalization group equations coupling $c_1$ to $c_2$,
$c_2$ to $c_1$ and $c_3$, and so forth. Generally speaking, for low 
values of $Q^2$, i.e.\ $Q^2 < 1$ GeV$^2$, and most hadrons, with the 
possible exception of pions, the valence quark state will be the 
largest term in the expansion, but the contributions from other 
terms may not be negligible.

This goes hand in hand with a change of the effective quark mass. 
While the assumption of quasi free partons with current quark masses
less than 10 MeV for light flavors is valid for $Q^2 > 1\ldots 2$ GeV$^2$, we 
know from lattice calculations and other considerations that the 
dynamical quark mass $m(Q^2)$ increases rapidly for $Q^2 < 1$ GeV$^2$ 
and approaches a constituent mass of order 300 MeV \cite{Bowman:04,
DiaPe:84,Roberts:00}.

For the sudden recombination of medium constituents into a fast
hadron, the relevant scale is determined by the average thermal 
momenta in the medium. The precise value of $Q^2$ depends on the
process that is being considered, but generally the scale falls
into the range $(\pi T)^2 \leq Q^2 \leq (2\pi T)^2$, where $T$
is the temperature of the medium \cite{HuaLis:94}. For a quark-gluon
plasma near the point of hadronization ($T_c \approx 170$ MeV), this
implies that the relevant scale $Q$ is in the range $0.5\ldots 1$ GeV.
It is therefore natural to assume that the correct degrees of freedom
for recombination act like constituent quarks and that the 
recombination probability is dominated by the lowest Fock states.

Nevertheless, even for very massive constituents, higher Fock states
should be present, even though we can not calculate their contribution from 
first principles. We want to discuss in the following, how an admixture
of higher Fock states alters the recombination formalism.

In the Boltzmann approximation, the probability for finding a quark 
(or gluon) with momentum $\mathbf k = x\mathbf P$ and energy 
$E_{\mathbf k}$ in the medium is given by:
\begin{equation}
w_q(x) = \langle q(x)|\hat\rho| q(x)\rangle
        = e^{-E_{\mathbf k}/T} = e^{-xP/T}
\label{eq03}
\end{equation}
where 
$\hat\rho$ denotes
the thermal density matrix and masses are assumed to be much smaller than
$Pc$. Hence for a state with $n$ quarks or other partons we have
\begin{eqnarray}
  &&\left\langle q(x_a){q}(x_b)\ldots \left| \hat\rho 
  \right|q(x_a){q}(x_b)\ldots \right\rangle \nonumber \\
  && = w_q(x_a) w_q(x_b)\ldots = e^{-(x_a+x_b+\ldots) P/T}
\end{eqnarray}
The emission probability for the single Fock space
components of a meson then is
\begin{widetext}
\begin{eqnarray}
W_{q\bar q} &=& \int_0^1 dx_a dx_b \delta(x_a+x_b-1) |c_1(x_a,x_b)|^2 
    \left\langle q(x_a){\bar q}(x_b)\left| \hat\rho 
    \right|q(x_a){\bar q}(x_b)\right\rangle 
    = C_1 e^{-P/T}, 
\label{eq04}
\\
W_{q{\bar q}g} &=& \int_0^1 dx_a dx_b dx_c \delta(x_a+x_b+x_c-1) 
    |c_2(x_a,x_b,x_c)|^2 \left\langle q(x_a){\bar q}(x_b)g(x_c)\left| 
    \hat\rho \right|q(x_a){\bar q}(x_b)g(x_c)\right\rangle
    = C_2 e^{-P/T},
\label{eq05}
\end{eqnarray}
\end{widetext}
etc. The delta functions expressing the condition that the sum of the 
light-cone momenta of all constituents add up to the light-cone 
momentum of the hadron ensure that the product of the Boltzmann 
factors of all constituents combine to the common factor $e^{-P/T}$
containing only the hadron momentum. It is obvious that the same
argument holds for any arbitrary complex Fock space component in
the hadron wave function, as long as the medium is a thermal one.
Note that we have suppressed quark flavor indices in the notation and factors 
arising from recombination of color, spin, and flavor quantum numbers, 
which can be dealt with similarly but at the expense of a greatly 
complicated notation. 

Combining the contributions from all Fock space components, the 
probability for emission of a hadron with momentum $P$ is given by:
\begin{equation}
W(P) = W_{q\bar q} + W_{q{\bar q}g} + W_{q{\bar q}q{\bar q}} + \cdots
 = e^{-P/T}
\label{eq06}
\end{equation}
where we applied the normalization condition (\ref{eq02}). 
Our result shows that the probability of relativistic emission of
a complex state by recombination from a thermal ensemble does not
depend on the degree of complexity of the state. Per spin-flavor
degree of freedom the emission of a baryon with momentum $P$ is
as likely as the emission of a meson with the same momentum, as
long as the particle masses are negligible compared with $P=|\mathbf P|$.
We call this property the ``egalitarian'' nature of recombination
from a thermal ensemble.

Let us next explore how the elliptic flow of hadrons is affected
by the presence of higher Fock states in their wavefunction. Here
we assume that there are no space-momentum correlations affecting
the calculation \cite{Molnar:04,PrattPal:04} and we limit our discussion
to sufficiently small values of the elliptic flow parameter $v_2$,
so that nonlinear corrections to the additivity rule of the flow
of the constituents can be safely neglected \cite{F03prc}. The 
expression for
the hadronic elliptic flow $v_2^{\rm (H)}(P)$ in terms of the 
quark/gluon elliptic flow $v_2(k)$ is:
\begin{eqnarray}
v_2^{\rm (H)}(P) &=& \sum_\nu \int_0^1 \left(\prod_\alpha dx_\alpha\right)
    \delta\left(\sum_\alpha x_\alpha -1\right) 
\nonumber \\
&&  \left(\sum_\alpha v_2(x_\alpha P)\right) \left|c_\nu(x_\alpha)\right|^2,
\label{eq07}
\end{eqnarray}
where $\nu$ denotes the different Fock space configurations and 
$\alpha$ enumerates the constituents of the hadron in each 
configuration. 
Here it is important to be in a light cone frame where the masses 
of the particles can be neglected.

In the hydrodynamic regime, where $v_2(k) = a k$,
one finds again trivially $v_2^{\rm (H)}(P) = a P$ using (\ref{eq02}). 
The elliptic flow of all hadrons then follows the same universal line.
In the saturation regime, the relationship is more complicated and 
one may suspect that higher Fock states alter the $v_2$ of the hadron.

For the derivation of the familiar scaling law for elliptic flow it is
assumed that the wave function of the hadron is narrow: all partons in 
a Fock state carry roughly equal momentum $x_i \approx 1/n_\nu$, where 
$n_\nu$ is the number of partons. The scaling law follows when only 
the lowest Fock state with $n_1$ partons is taken into account 
\cite{Vo02,F03prc}
\begin{equation}
  v_2^{\rm (H)}(P) = n_1 v_2 (P/n_1)
\label{eq08a}
\end{equation}
The experimental data is described very well by this equation
\cite{Sorensen:2003wi,F03prc}.
We can easily generalize (\ref{eq08a}) to higher Fock states in the limit
of a very narrow wave function $\delta(\sum_i x_i -1) |c_\nu (x_\alpha)|^2 
\approx C_\nu \prod_i \delta(x_i - 1/n_\nu)$.
Then
\begin{equation}
  v_2^{\rm (H)}(P) \approx \sum_\nu C_\nu n_\nu v_2(P/n_\nu)
\label{eq08}
\end{equation}
and the scaling law is apparently violated by the contributions from
higher Fock states. In principle, this violation should be visible in 
a scaling analysis. The data are usually plotted with scaled axes 
$P_T/n_1$ and $v_2/n_1$, where $n_1=2,3$ is the valence quark number 
for the hadron. Equation (\ref{eq08}) implies that the scaled elliptic 
flow for mesons and baryons, respectively, is
\begin{eqnarray}
  \tilde v_2^{\rm (M)}(p) &=& 
  \sum_\nu C_\nu^{\rm (M)} \frac{n_\nu^{\rm (M)}}{2} v_2 \left(2p/
  n_\nu^{\rm (M)}\right) 
    \nonumber \\
  \tilde v_2^{\rm (B)}(p) &=& 
  \sum_\nu C_\nu^{\rm (B)} \frac{n_\nu^{\rm (B)}}{3} v_2 \left(3p/
  n_\nu^{\rm (B)}\right) .
\label{eq10}
\end{eqnarray}
Clearly, if all $C_\nu=0$ except the lowest Fock states, for which 
$C_1^{\rm (M)}=1$ and $C_1^{\rm (B)}=1$, the scaled elliptic 
flow curve is the same for mesons and baryons. This is what has been found 
in the data and has been interpreted as evidence that the scaled curve 
reflects the partonic elliptic flow before hadronization: 
$\tilde v^{\rm (M)}(p) = \tilde v^{\rm (B)}(p) = v_2(p)$.

Can higher Fock states be present despite of the scaling law holding?
We notice that for states with the same number $s$ of sea quarks or gluons
the ratio of prefactors $n_\nu/n_1$ for mesons and baryons in (\ref{eq10}) is
\begin{equation}
  \frac{3}{2}\frac{s+2}{s+3}
\end{equation}
and the ratio of prefactors inside the argument, multiplying the
momentum $p$ is the inverse of that.
This ratio is always between 1 for the pure valence state and 3/2 for 
states with an infinite number of sea quarks and gluons. This is not a
large variation, and there are two other effects that are also important. 
First, we note that the prefactors in front of $v_2$ and the momentum $p$
in each line of (\ref{eq10}) tend to cancel if $v_2$ is a rising function.
Second, with increasing number of sea quarks and gluons the average
momentum of each parton decreases. For large Fock states we expect the
partons to be in the hydrodynamic regime where $v_2$ is a linear function
and scaling holds for arbitrary parton numbers.

How large can the scaling violation be expected to be in practice? 
In order to present a numerical estimate, we consider the lowest Fock
state and the next higher one, with an additional gluon, for mesons and 
baryons respectively. In focusing on the one-gluon admixture, we assume
that the weights of higher Fock states, including two or more additional 
gluons or one or more quark-antiquark pairs, die out rapidly with growing
complexity. For this analysis we abandon the simplistic $\delta$-function 
shaped wave functions. We use the following more realistic light-cone 
distribution amplitudes for mesons as guidance:
\begin{eqnarray}
\frac{c_1^{\rm (M)}(x_a,x_b)}{\sqrt{x_a x_b}} 
  &=& \sqrt{C_1^{\rm (M)}}\frac{x_a x_b}{2\sqrt{35}} ,
\nonumber \\
\frac{c_2^{\rm (M)}(x_a,x_b,x_g)}{\sqrt{x_a x_b x_g}} 
  &=& \sqrt{C_2^{\rm (M)}}\frac{x_a x_b x_g^2}{12\sqrt{10010}} ;
\label{eq11}
\end{eqnarray}
and for baryons:
\begin{eqnarray}
\frac{c_1^{\rm (B)}(x_a,x_b,x_c)}{\sqrt{x_a x_b x_b}}
  &=& \sqrt{C_1^{\rm (B)}}\frac{x_a x_b x_c}{4\sqrt{330}} ,
\nonumber \\
\frac{c_2^{\rm (B)}(x_a,x_b,x_c,x_g)}{\sqrt{x_a x_b x_c x_g}}
  &=& \sqrt{C_2^{\rm (B)}}\frac{x_a x_b x_c x_g^2}{1680\sqrt{4862}} .
\label{eq12}
\end{eqnarray}
Here the variable $x_g$ denotes the light-cone momentum
fraction of the additional gluon in the hadron wave function. The
model wave functions on the right-hand side of (\ref{eq11}) and (\ref{eq12}) 
are standard forms of higher twist distributions for hadrons \cite{Ba98}. The
additional kinematic factors on the left-hand side arise because
of our non-standard probabilistic normalization of the states.

\begin{figure}
  \begin{center}
  \epsfig{file=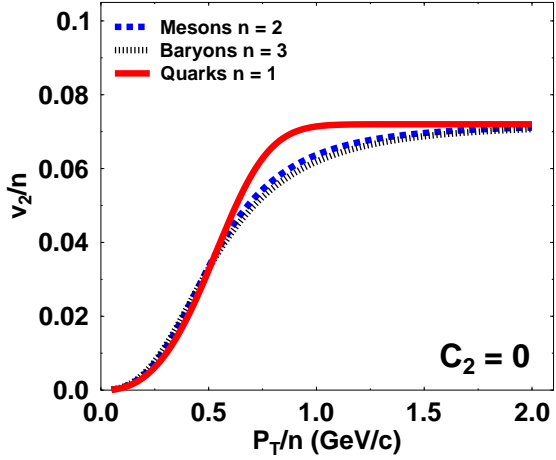,width=0.9\columnwidth}
  \end{center}
  \caption{\label{fig1} Scaled elliptic flow of mesons and 
   baryons (dotted lines) compared to  the quark flow (solid line).
   The difference between the quark and scaled hadron flow is due to
   the smearing by the internal hadron wavefunctions.}
\end{figure}

\begin{figure}
  \begin{center}
  \epsfig{file=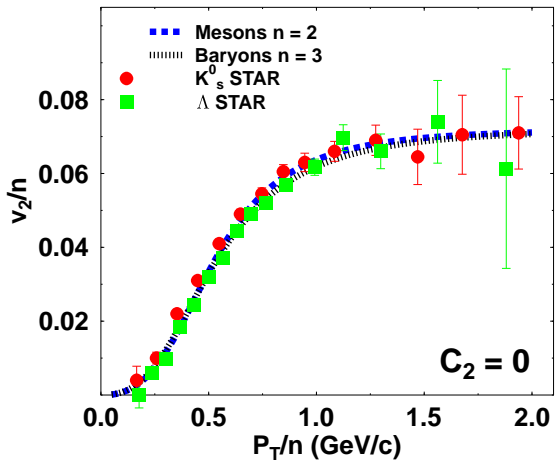,width=0.9\columnwidth}
  \end{center}
  \caption{\label{fig2} Scaled elliptic flow of baryons and mesons as 
   calculated from the quark flow compared to data for 
   $\Lambda$-hyperons and kaons.}
\end{figure}

\begin{figure}
  \begin{center}
  \epsfig{file=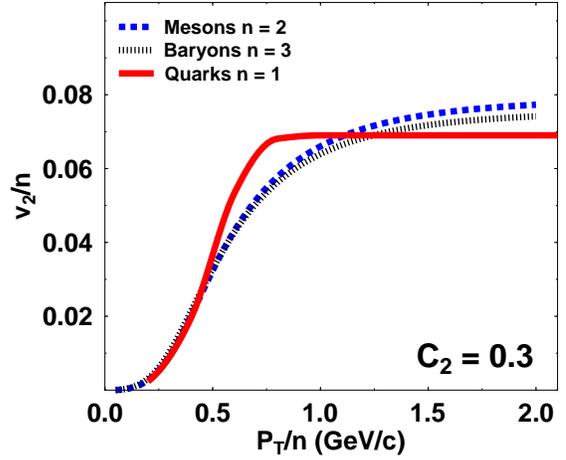,width=0.9\columnwidth}
  \end{center}
  \caption{\label{fig3} Scaled elliptic flow of mesons and baryons 
   (dotted lines) compared to  the parton flow (solid line), but now
   for hadron wavefunctions with a 30 percent component containing an 
   additional gluon.}
\end{figure}

\begin{figure}
  \begin{center}
  \epsfig{file=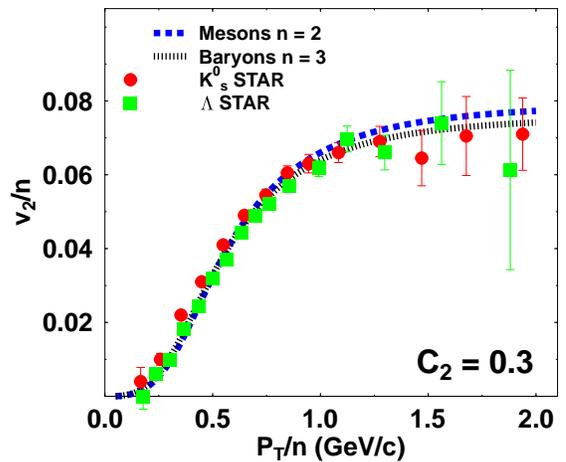,width=0.9\columnwidth}
  \end{center}
  \caption{\label{fig4} Scaled elliptic flow of baryons and mesons as 
   calculated from the parton flow, utilizing hadron wavefunctions with 
   a 30 percent component containing an additional gluon, compared to 
   data for $\Lambda$-hyperons and kaons.}
\end{figure}

The constants $C_i$ give the relative normalization of the two
Fock states.
In the following, we shall consider two cases: ({\em i}) lowest Fock state 
only ($C_1=1, C_2=0$) and ({\em ii}) moderate $|q{\bar q}g\rangle$ or 
$|qqqg\rangle$contribution ($C_1=0.7, C_2=0.3$).
It is important to recognize that even in the first case, although
valence quark scaling is well realized, the scaled elliptic flow 
function $\tilde v_2^{\rm (H)}$ of the hadrons differs from the input 
function $v_2$ describing the quark flow. This is illustrated in
Fig.~\ref{fig1}, which shows the scaled elliptic flow of mesons and 
baryons (dotted lines) together with the quark flow (solid line).
The difference between the parton and scaled hadron flow is due to
the smearing by the internal hadron wave functions. Figure \ref{fig2}
shows the scaled hadron elliptic flow in comparison with the scaled 
STAR data \cite{Sorensen:2003wi} for neutral kaons and $\Lambda$ 
hyperons. The quark flow function was chosen so that the theoretical 
values obtained by use of Eq.~(\ref{eq07}) fit the data. The figure 
demonstrates that a single input curve for quarks can describe 
the elliptic flow of both, mesons and baryons.

Figures \ref{fig3} and \ref{fig4} show the same quantities for the case
({\em ii}), i.~e.\ for hadron wave functions with a 30 percent component 
containing an additional gluon. We assume that the elliptic
flow for the gluons is the same as that for light quarks.
Note that the parton elliptic flow has been slightly adjusted to yield a 
good fit to the data again for this case. 
As anticipated, the scaling is not as good as for the lowest Fock state 
alone but, as Fig.~\ref{fig3} demonstrates, the scaling is only modestly 
violated. The largest violation occurs for high transverse momenta. 
However, we also expect deviations to start in this region
because fragmentation starts to replace recombination as the dominate 
hadronization mechanism. The present precision of the data, as seen 
in Fig.~\ref{fig4}, is insufficient to observe the deviations from 
the scaling law and a gluon contribution of 30\% is compatible with
the data. 

\begin{figure}
  \begin{center}
  \epsfig{file=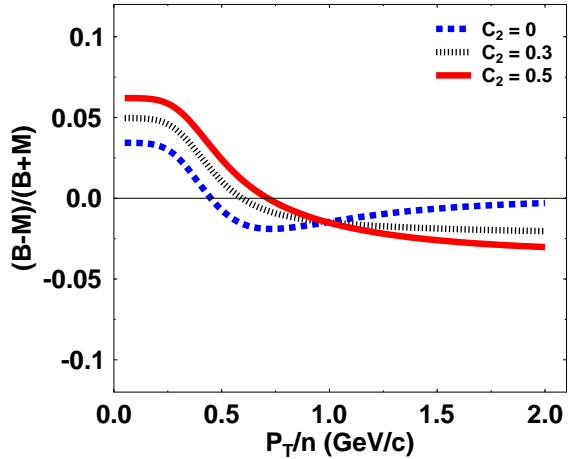,width=0.9\columnwidth}
  \end{center}
  \caption{\label{fig5} Scaled elliptic flow of mesons and 
   baryons (dotted lines) compared to  the quark flow (solid line).
   The difference between the quark and scaled hadron flow is due to
   the smearing by the internal hadron wavefunctions.}
\end{figure}

Since the parton elliptic flow can be adjusted after including higher
Fock states, the real test is the systematic deviation between 
baryons and mesons.
In Figure \ref{fig5} we show the relative difference $(\tilde v_2^{\rm (B)}
-\tilde v_2^{\rm (M)})/(\tilde v_2^{\rm (B)}+\tilde v_2^{\rm (M)})$
between the scaled meson and baryon elliptic flow for three different 
sizes of the higher Fock state component (0\%, 30\%, 50\%). In all cases, 
baryons have a slightly larger scaled $\tilde v_2$ than mesons at small 
momenta. This effect is likely to be overwhelmed by the influence of mass 
differences, which have been neglected in the sudden recombination model. 
At larger momenta, the scaled meson $\tilde v_2$ is slightly larger. As 
stated, these differences are well within present experimental errors.

In summary, we have investigated the effects of higher Fock states on 
the recombination of hadrons from thermalized quark distributions.
We showed that the yield of relativistic parton clusters is independent of 
the number of partons in the cluster. Therefore, hadron spectra remain 
unaffected by the inclusion of higher Fock states. One important 
implication is that gluon degrees of freedom could be accomodated
during hadronization. They simply become part of the 
quark-gluon wave functions of the produced hadrons, but remain
hidden constituents because the commonly produced hadrons do not
contain valence gluons.

On the other hand higher Fock states introduce deviations from the
scaling law for elliptic flow. We showed that an additional 30\% 
contribution from gluons is compatible with the existing data on
elliptic flow from RHIC.
We emphasize that the interpretation that elliptic flow data from RHIC
proves the existence of quark degrees of freedom in the bulk matter 
produced in the heavy ion collision is still valid. However, the
connection of the measured elliptic flow to the quark elliptic flow
might be less straight forward than anticipated.

{\em Acknowledgments:} We thank A.~Sch\"afer for critical comments
which motivated this work. This work was supported in part by the
DOE grants DE-FG02-96ER40945 and DE-FG02-87ER40328.


\begin{thebibliography}{99}

\bibitem{phenix:03ppi}
K.~Adcox {\it et al.}  (PHENIX Collaboration),
{Phys.\ Rev.\ Lett.} {\bf 88}, 242301 (2002);
S.~S.~Adler {\it et al.} (PHENIX Collaboration),
{Phys.\ Rev.\ Lett.} {\bf 91} 172301 (2003);
S.~S.~Adler {\it et al.} (PHENIX Collaboration),
{Phys.\ Rev.} C {\bf 69} 034909 (2004).

\bibitem{Velkovska:2004cb}
J.~Velkovska,
J.\ Phys.\ G {\bf 30}, S151 (2004).

\bibitem{v2_data}
S.~S.~Adler {\it et al.}  [PHENIX Collaboration],
Phys.\ Rev.\ Lett.\  {\bf 91}, 182301 (2003);
J.~Adams {\it et al.}  [STAR Collaboration],
Phys.\ Rev.\ Lett.\  {\bf 92}, 052302 (2004).

\bibitem{Sorensen:2003wi}
P.~Sorensen  [STAR Collaboration],
J.\ Phys.\ G {\bf 30}, S217 (2004)
[arXiv:nucl-ex/0305008].

\bibitem{Vo02}
S.~A.~Voloshin,
Nucl.\ Phys.\ A {\bf 715}, 379 (2003);

\bibitem{Fr03a}
R.~J.~Fries, B.~M\"uller, C.~Nonaka and S.~A.~Bass,
Phys.\ Rev.\ Lett.\  {\bf 90}, 202303 (2003);
J.\ Phys.\ G {\bf 30}, 223 (2004).

\bibitem{F03prc}
R.~J.~Fries, B.~M\"uller, C.~Nonaka and S.~A.~Bass,
Phys.\ Rev.\ C {\bf 68}, 044902 (2003).

\bibitem{Fr04}
R.~J.~Fries,
J.\ Phys.\ G {\bf 30}, S853 (2004).

\bibitem{Gr03a}
V.~Greco, C.~M.~Ko and P.~Levai,
Phys.\ Rev.\ Lett.\  {\bf 90}, 202302 (2003);
Phys.\ Rev.\ C {\bf 68}, 034904 (2003).

\bibitem{MV03}
D.~Molnar and S.~A.~Voloshin,
Phys.\ Rev.\ Lett.\  {\bf 91}, 092301 (2003).

\bibitem{Hw03a}
R.~C.~Hwa and C.~B.~Yang,
Phys.\ Rev.\ C {\bf 67}, 034902 (2003);
Phys.\ Rev.\ C {\bf 67}, 064902 (2003).

\bibitem{Bowman:04}
P.~O.~Bowman, U.~M.~Heller, D.~B.~Leinweber, A.~G.~Williams and J.~B.~Zhang,
{\it Nucl.\ Phys.\ Proc.\ Suppl.}  {\bf 128}, 23 (2004).

\bibitem{DiaPe:84}
D.~Diakonov and V.~Y.~Petrov,
{\it Phys.\ Lett.} B {\bf 147}, 351 (1984);
{\it Nucl.\ Phys.} B {\bf 272}, 457 (1986);
D.~Diakonov,
arXiv:hep-ph/0406043.

\bibitem{Roberts:00}
C.~D.~Roberts and S.~M.~Schmidt,
{\it Prog.\ Part.\ Nucl.\ Phys.}  {\bf 45}, S1 (2000).

\bibitem{HuaLis:94}
S.~Z.~Huang and M.~Lissia,
Nucl.\ Phys.\ B {\bf 438}, 54 (1995).

\bibitem{Molnar:04}
D.~Molnar,
arXiv:nucl-th/0408044.

\bibitem{PrattPal:04}
S.~Pratt and S.~Pal,
arXiv:nucl-th/0409038.

\bibitem{Ba98}
V.~M.~Braun and I.~E.~Filyanov,
Z.\ Phys.\ C {\bf 48}, 239 (1990);
P.~Ball, V.~M.~Braun, Y.~Koike and K.~Tanaka,
Nucl.\ Phys.\ B {\bf 529}, 323 (1998).

\end{thebibliography}
\end{document}